\input{aipcheck}

\documentclass[
    ,final            % use final for the camera ready runs
  ]
  {aipproc}

\layoutstyle{6x9}

\def\aa{{\em A\&A}\ }

\def\aj{{\em AJ}\ }
\def\annrev{{\em ARA\&A}\ }
\def\apj{{\em ApJ}\ }
\def\apjs{{\em ApJS}\ }

\def\nat{{\em Nature}\ }

\def\lsim{\mathrel{\rlap{\lower 4pt \hbox{\hskip 1pt $\sim$}}\raise 1pt
\hbox {$<$}}} 
\def\gsim{\mathrel{\rlap{\lower 4pt \hbox{\hskip 1pt $\sim$}}\raise 1pt
\hbox {$>$}}}
\newcommand{\etal}{et~al.}

\newcommand{\eg}{e.g., }

\newcommand{\Msun}{M_{\odot}}

\newcommand{\kms}{km~s$^{-1}$}

\newcommand{\OI}{O~{\sc i}}
\newcommand{\SiII}{Si~{\sc ii}}
\newcommand{\CaII}{Ca~{\sc ii}}
\newcommand{\FeII}{Fe~{\sc ii}}
\newcommand{\FeIII}{Fe~{\sc iii}}
\newcommand{\TiII}{Ti~{\sc ii}}

\newcommand{\Nifs}{$^{56}$Ni}
\newcommand{\Mej}{M_{\rm ej}}

\newcommand{\Mni}{M{\rm (^{56}Ni)}}
\newcommand{\Ed}{\dot{E}_{\rm dep}}

\newcommand{\Edep}{\dot{E}_{\rm dep,51}}

\begin{document}

\title{Nucleosynthesis in Core-Collapse Supernovae and 
 GRB--Metal-Poor Star Connection}


\author{K.~Nomoto}{
  address={Department of Astronomy, University of Tokyo, Bunkyo-ku, Tokyo
113-0033, Japan}
}

\author{N.~Tominaga}{
  address={Department of Astronomy, University of Tokyo, Bunkyo-ku, Tokyo
113-0033, Japan}
}

\author{M.~Tanaka}{
  address={Department of Astronomy, University of Tokyo, Bunkyo-ku, Tokyo
113-0033, Japan}
}

\author{K.~Maeda}{
  address={Max-Planck-Institut f\"ur Astrophysik, 
85741 Garching, Germany}
}

\author{H.~Umeda}{
  address={Department of Astronomy, University of Tokyo, Bunkyo-ku, Tokyo
113-0033, Japan}
}

\begin{abstract}

We review the nucleosynthesis yields of core-collapse supernovae (SNe)
for various stellar masses, explosion energies, and metallicities.
Comparison with the abundance patterns of metal-poor stars provides
excellent opportunities to test the explosion models and their
nucleosynthesis.  We show that the abundance patterns of extremely
metal-poor (EMP) stars, e.g., the excess of C, Co, Zn relative to Fe,
are in better agreement with the yields of hyper-energetic explosions
(Hypernovae, HNe) rather than normal supernovae.

We note that the variation of the abundance patterns of EMP stars are
related to the diversity of the Supernova-GRB connection.  We
summarize the diverse properties of (1) GRB-SNe, (2) Non-GRB HNe/SNe,
(3) XRF-SN, and (4) Non-SN GRB.  In particular, the Non-SN GRBs (dark
hypernovae) have been predicted in order to explain the origin of
C-rich EMP stars.  We show that these variations and the connection
can be modeled in a unified manner with the explosions induced by
relativistic jets.
Finally, we examine whether the most luminous supernova 2006gy can be
consistently explained with the pair-instability supernova model.

\end{abstract}

\maketitle

\begin{center}
{\small To appear in "Supernova 1987A: 20 Years After: Supernovae and
Gamma-Ray Bursters", eds. S. Immler, K. Weiler, \& R. McCray (American Institute of Physics) (2007) }
\end{center}

%%%%%%%%%%%%%%%%%%%%%%%%%%%%%%%%%%%%%%%%%%%%
%% MAINMATTER
%%%%%%%%%%%%%%%%%%%%%%%%%%%%%%%%%%%%%%%%%%%%

\section{Introduction}

Massive stars in the range of 8 to $\sim$ 130$M_\odot$ undergo
core-collapse at the end of their evolution and become Type II and
Ib/c supernovae (SNe) unless the entire star collapses into a black
hole with no mass ejection \cite[e.g.,][]{arnett1996, hillebrandt2003,
fryer2004}.

The explosion energies of core-collapse supernovae are fundamentally
important quantities, and an estimate of $E \sim 1\times 10^{51}$ ergs
has often been used in calculating nucleosynthesis and the impact on
the interstellar medium.  (In the present paper, we use the explosion
energy $E$ for the final kinetic energy of explosion, and $E_{51} =
E/10^{51}$\,erg.)  A good example is SN1987A in the Large Magellanic
Cloud, whose energy is estimated to be $E_{51} = 1.0 - 1.5$ from its
early light curve \cite[e.g.,][]{arnett1996, nomoto1994b}.

One of the most interesting recent developments in the study of
supernovae is the discovery of some very energetic supernovae, whose
kinetic energy (KE) exceeds $10^{52}$\,erg, more than 10 times the KE
of normal core-collapse SNe.  The most luminous and powerful of these
objects, the Type Ic supernova (SN~Ic) 1998bw, was linked to the
gamma-ray burst GRB 980425 \cite{galama1998}, thus establishing for
the first time a connection between long-duration gamma-ray bursts
(GRBs) and the well-studied phenomenon of core-collapse SNe
\cite{woo06}.  However, SN~1998bw was exceptional for a SN~Ic: it was
as luminous at peak as a SN~Ia, indicating that it synthesized $\sim
0.5 M_\odot$ of $^{56}$Ni, and its KE was estimated at $E_{51} \sim
30$ \cite{iwa98}.

In the present paper, we use the term 'Hypernova (HN)' to describe
such a hyper-energetic supernova with $E \gsim 10^{52}$ ergs without
specifying the explosion mechanism \cite{nomoto2001}.  Following SN
1998bw, other ``hypernovae'' of Type Ic have been discovered or
recognized \cite{nom04}.  

\begin{figure*}
\includegraphics*[width=10cm]{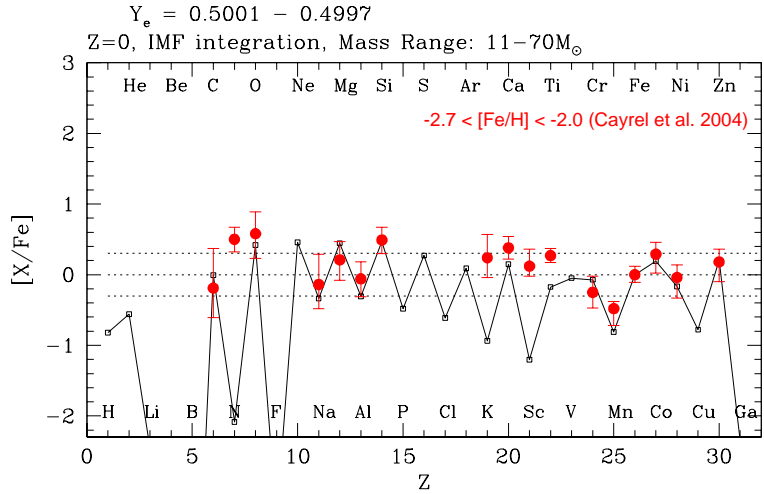}
\caption{Comparison between the abundance pattern of VMP stars
\cite{cayrel2004} ({\it filled circles with error bars}) and the IMF
integrated yield of Pop III SNe from 10$\Msun$ to 50 $\Msun$
\cite{tominaga2006}
}
\label{fig:IMF}
\end{figure*}

Nucleosynthesis features in such hyper-energetic (and
hyper-aspherical) supernovae must show some important differences from
normal supernova explosions.  This might be related to the unpredicted
abundance patterns observed in the extremely metal-poor (EMP) halo
stars \cite[e.g.,][]{hill2005, beers2005}.  This approach leads to
identifying the First Stars in the Universe, i.e., metal-free,
Population III (Pop III) stars which were born in a primordial
hydrogen-helium gas cloud.  This is one of the important challenges of
the current astronomy.

\section{Abundance Patters of Metal-Poor Stars}

We have calculated the nucleosynthesis yields for various stellar
masses, explosion energies, and metallicities
\cite{nomoto2006,koba06,tominaga2006}.  From the light curve and
spectra fitting of individual supernova, the relations between the
mass of the progenitor, explosion energy, and produced $^{56}$Ni mass
have been obtained.

The enrichment by a single SN can dominate the preexisting metal
contents in the early universe.  Therefore, the comparison between the
SN model and the abundance patterns of EMP stars can provide a new way
to find out the individual SN nucleosynthesis.

\subsubsection{Very Metal-Poor (VMP) Stars}

VMP stars defined as [Fe/H] $\lsim -2.5$ \cite{beers2005} are likely
to have the abundance pattern of well-mixed ejecta of many SNe.  We
thus compare the abundance patters of VMP stars with the SN yields
integrated over the progenitors of 10 - 50 $M_\odot$
(Fig.~\ref{fig:IMF}).

Since the abundance patterns of supernova models with [Fe/H] $\lsim
-2.5$ are quite similar to those of Pop III star models
\cite{umeda2000, woo95}, we use the Pop III yields for VMP and EMP
stars.  Comparison between the integrated yields over the Salpeter's
IMF and the abundance pattern of VMP stars (Fig.~\ref{fig:IMF}) show
that many elements are in reasonable agreements (see \cite{nomoto2006}
for further details).

\subsubsection{Extremely Metal-Poor (EMP) Stars}

\begin{figure*}
\includegraphics*[width=10cm]{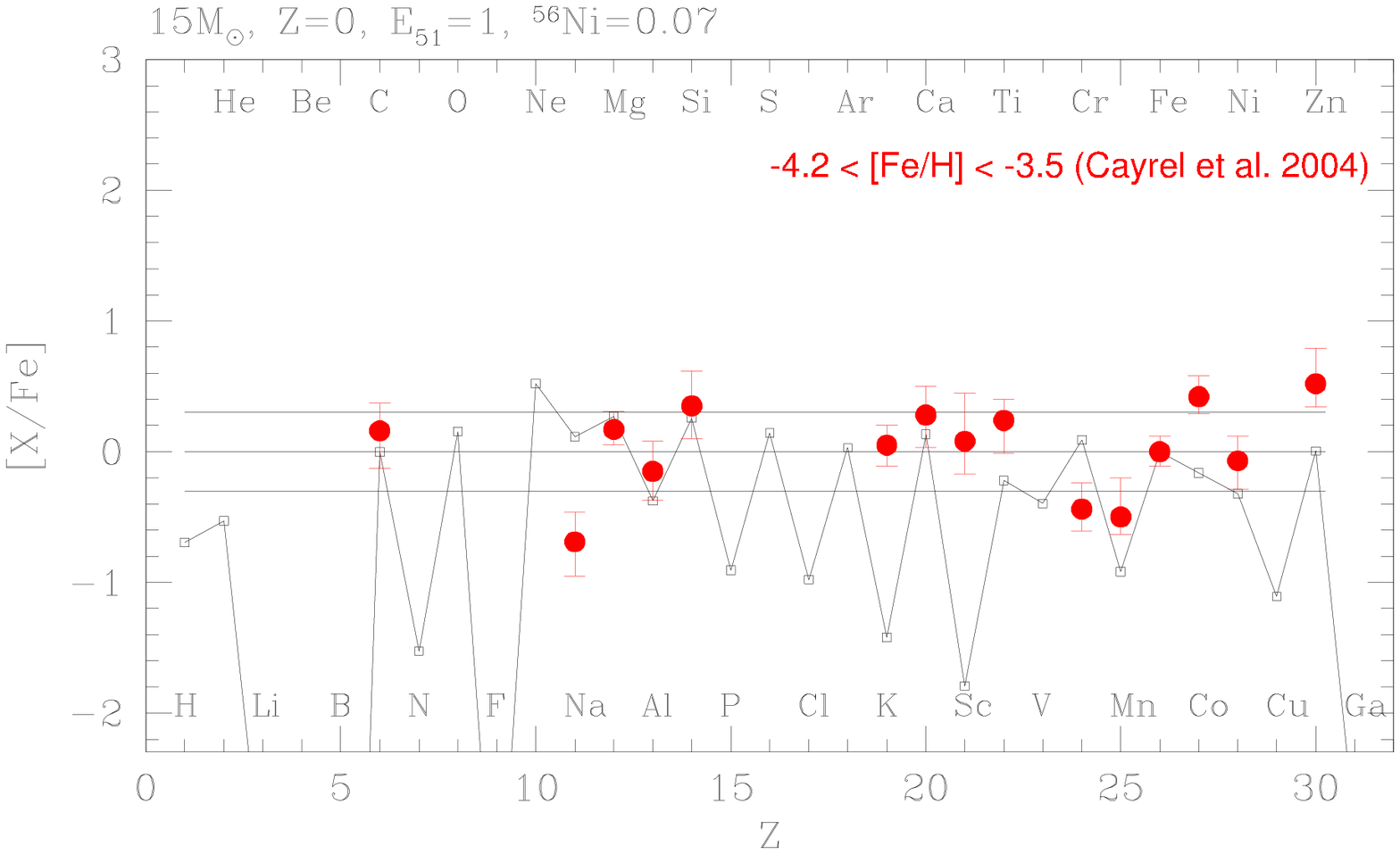}
\caption{Averaged elemental abundances of stars with [Fe/H] $= -3.7$
\cite{cayrel2004} compared with the normal SN yield (15
$M_\odot$, $E_{51} =$ 1).
}
\label{fig7}
\end{figure*}

\begin{figure*}
\includegraphics*[width=10cm]{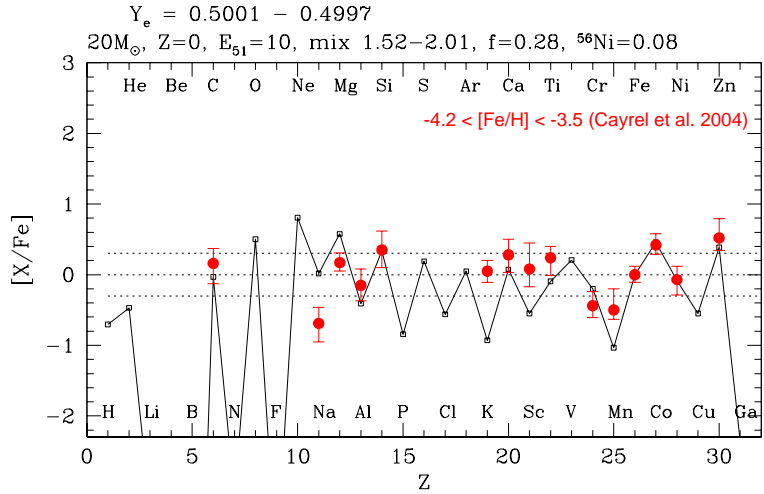}
\caption{Averaged elemental abundances of stars with [Fe/H] $= -3.7$
\cite{cayrel2004} compared with the hypernova yield (20 $M_\odot$,
$E_{51} =$ 10).}
\label{fig7h}
\end{figure*}

In the early galactic epoch when the galaxy was not yet chemically
well-mixed, each EMP star ([Fe/H] $\lsim -2.5$) may be formed mainly
from the ejecta of a single Pop III SN (although some of them might be
the second or later generation SNe) \cite[e.g.,][]{argast2000, tum06}.
The formation of EMP stars was driven by a supernova shock, so that
[Fe/H] was determined by the ejected Fe mass and the amount of
circumstellar hydrogen swept-up by the shock wave.  Then, hypernovae
with larger $E$ are likely to induce the formation of stars with
smaller [Fe/H], because the mass of interstellar hydrogen swept up by
a hypernova is roughly proportional to $E$ \cite{thorn98} and the
ratio of the ejected iron mass to $E$ is smaller for hypernovae than
for normal supernovae.

The theoretical yields are compared with the averaged abundance
pattern of four EMP stars, CS~22189-009, CD-38:245, CS~22172-002 and
CS~22885-096, which have low metallicity ($-4.2<{\rm [Fe/H]}<-3.5$)
and normal [C/Fe] $\sim 0$ \cite{cayrel2004}.

Figures \ref{fig7} and \ref{fig7h} show that the averaged abundances
of EMP stars can be fitted well with the hypernova model of 20
$M_\odot$ and $E_{51} =$ 10 (Fig. \ref{fig7h}) but not with the normal
SN model of 15 $M_\odot$ and $E_{51} =$ 1 (Fig. \ref{fig7})
\cite{nomoto2005, tominaga2006}.

In the normal SN model (Fig. \ref{fig7}), the mass-cut is determined
to eject Fe of mass 0.14 $M_\odot$).  Then the yields are in
reasonable agreements with the observations for [(Na, Mg, Si)/Fe], but
give too small [(Mn, Co, Ni, Zn)/Fe] and too large [(Ca, Cr)/Fe].

In the HN model (Fig. \ref{fig7h}), these ratios are in much better
agreement with observations.  The ratios of Co/Fe and Zn/Fe are larger
in higher energy explosions since both Co and Zn are synthesized in
complete Si burning at high temperature region (see the next
subsection).  To account for the observations, materials synthesized
in a deeper complete Si-burning region should be ejected, but the
amount of Fe should be small.  This is realized in the mixing-fallback
models \cite{umeda2002a, ume05}.

\section{Supernova--Gamma-Ray Burst Connection}

\begin{figure}[t]
\includegraphics[width=7.cm]{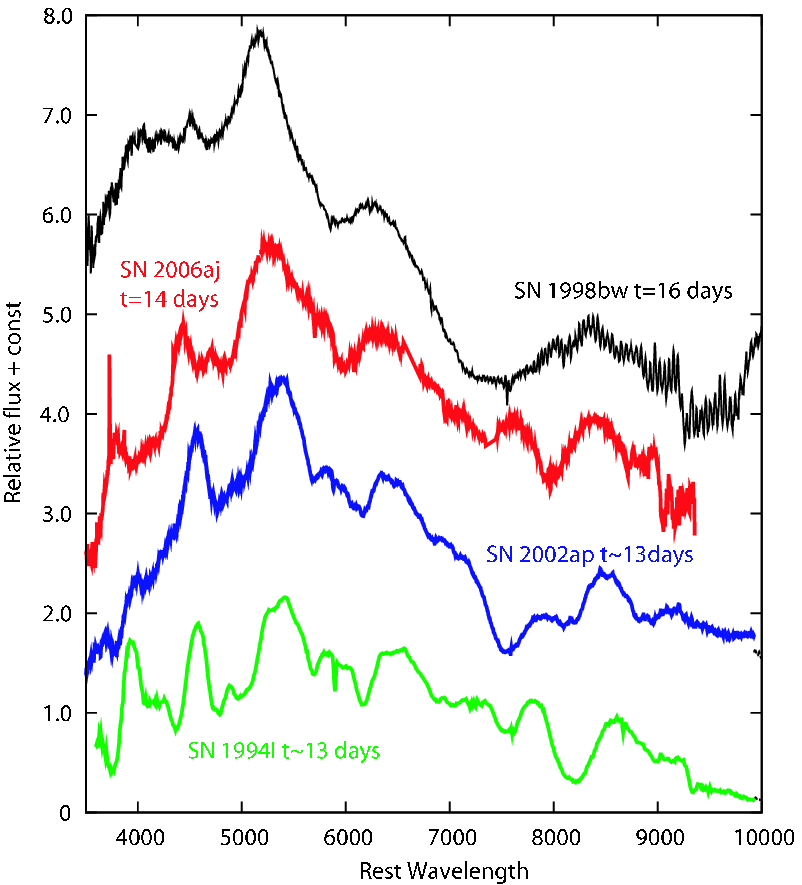}
\includegraphics[width=7.5cm]{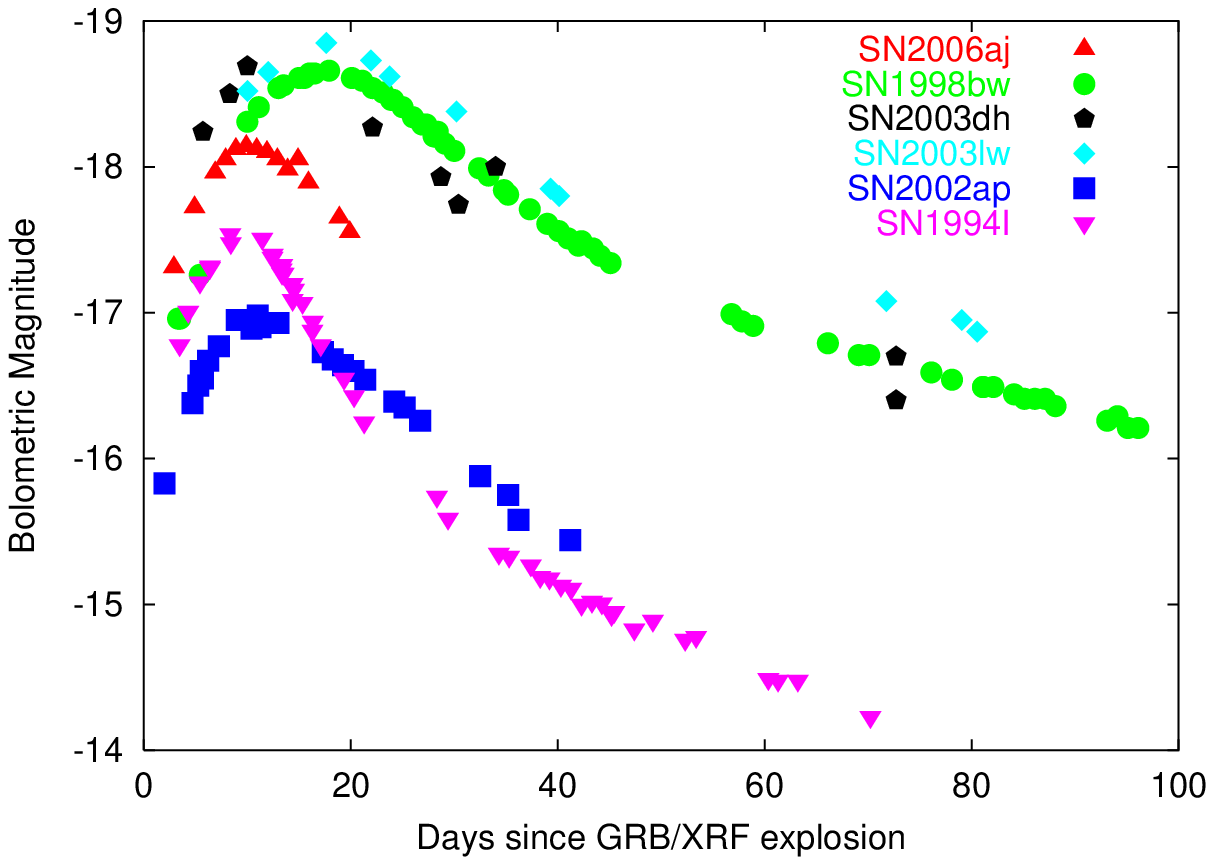}
\caption{(Left)
The spectra of 3 Hypernovae and 1 normal SN a few days before maximum.
SN~1998bw/GRB~980425 represents the GRB-SNe. SN~2002ap is a non-GRB
Hypernova.  SN~2006aj is associated with XRF~060218, being similar to
SN~2002ap. SN~1994I represents normal SNe.  (Right) The bolometric
light curves of GRB-SN (SNe 1998bw, 2003dh), non-GRB-SN (2002ap),
XRF-SN (2006aj), and normal SNe Ic (1994I) are compared.
}
\label{figSP}
\end{figure}

We have shown that nucleosynthesis in HNe is in better agreement with
the abundance pattern of EMP stars.  Thus it would be useful to
examine the GRB-SN connection in relation to the GRB-First Star
connection.

GRBs at sufficiently close distances ($z<0.2$) have been found to be
accompanied by luminous core-collapse SNe Ic (GRB~980425/SN~1998bw
\cite{galama1998}; GRB~030329/SN~2003dh \cite{sta03, hjo03};
GRB~031203/SN~2003lw \cite{mal04}).  Such GRB-SN connection is now
revealing quite a large diversity as follows.

(1) GRB-SNe: The three SNe Ic associated with the above GRBs have
similar properties; showing broader lines than normal SNe Ic
(Fig. \ref{figSP}: so-called broad-lined SNe \cite{woo06, mod07}).
These three GRB-SNe have been all found to be Hypernovae (HNe), i.e.,
very energetic supernovae, whose {\sl isotropic} kinetic energy (KE)
exceeds $10^{52}$\,erg, about 10 times the KE of normal core-collapse
SNe \cite{iwa98,nomoto2001,nom04}.

(2) Non-GRB HNe/SNe: These SNe show broad line features but are not
associated with GRBs (SN~1997ef \cite{iwa00}; SN~2002ap \cite{maz02};
SN 2003jd \cite{maz05}).  These are either less energetic than
GRB-SNe, or observed off-axis.

(3) XRF-SNe: X-Ray Flash (XRF) 060218 has been found to be connected
to SN Ic 2006aj \cite{campana2006, pian2006, sod06}.  The progenitor's
mass is estimated to be small enough to form a ``neutron star-making
SN'' \cite{mazzali2006b}.

(4) Non-SN GRBs: We have pointed out that nucleosynthesis in HNe can
explain some of the peculiar abundance patterns (such as the large
Zn/Fe and Co/Fe ratios) in extremely metal-poor stars, which have long
been mysteries.  In particular, we have predicted that the ``dark HN''
($=$ ``non-SN GRB'' $=$ long GRB with no SN) should exist and be
responsible for the formation of the carbon-rich extremely (and hyper)
metal-poor stars \cite{nom06b}.  The predicted ``non-SN GRBs'' have
been actually discovered (GRB 060605 and 060614)
\cite{fyn06,gal06,del06,geh06}.

\subsubsection{GRB-Supernova}

Figure \ref{figSP} compares the spectra of GRB-HNe (SN~1998bw),
non-GRB SN, XRF-SNe, and normal SN Ic.  The spectrum of SN~1998bw has
very broad lines.  The strongest absorptions are \TiII-\FeII\
(shortwards of $\sim 4000$\AA, \FeII-\FeIII\ (near 4500\AA), \SiII\
(near 5700\AA), and \OI-\CaII\ (between 7000 and 8000 \AA).  We
calculate the synthetic spectra for ejecta models of bare C+O stars
with various ejected mass $\Mej$ and $E$.  The large $E$/$\Mej$ is
required to reproduce the broad features.

The spectroscopic modelings are combined with the light curve (LC)
modeling to give the estimates of $\Mej$ and $E$.  The timescale of
the LC around maximum brightness reflects the timescale for optical
photons to diffuse \cite{arn82}.  For larger $\Mej$ and smaller $E$,
the LC peaks later and the LC width becomes broader because it is more
difficult for photons to escape.

 From the synthetic spectra and light curves, it was interpreted as
the explosion of a massive star, with $E_{51} \sim 30$ and $\Mej \sim
10 \Msun$\cite{iwa98}.  Also the very high luminosity of SN 1998bw
indicates that a large amount of \Nifs\ ($\sim 0.5 \Msun$) was
synthesized in the explosion.

The ejected \Nifs\ mass is estimated to be $\Mni\sim0.3-0.7\Msun$ (\eg
\cite{maz06a}) which is 4 to 10 times larger than typical SNe Ic
($\Mni\sim 0.07\Msun$ \cite{nom94}).

The other two GRB-SNe, 2003dh and 2003lw, are also characterized by
the very broad line features and the very high luminosity.  $\Mej$ and
$E$ are estimated from synthetic spectra and light curves and
summarized in Figure \ref{figME}\cite{nak01a, deng05, maz06a}.  It is
clearly seen that GRB-SNe are the explosions of massive progenitor
stars (with the main sequence mass of $M_{\rm ms} \sim 35 - 50
\Msun$), have large explosion kinetic energies ($E_{51} \sim 30 -
50$), synthesized large amounts of \Nifs\ ($\sim 0.3 - 0.5 \Msun$).

These GRB-associated HNe (GRB-HNe) are suggested to be the outcome
of very energetic black hole (BH) forming explosions of massive stars
(\eg \cite{iwa98}).

\subsubsection{Non-GRB Hypernovae}

These HNe show spectral features similar to those of GRB-SNe but are
not known to have been accompanied by a GRB.  The estimated $\Mej$ and
$E$, obtained from synthetic light curves and spectra, show that there
is a tendency for non-GRB HNe to have smaller $\Mej$ and $E$, and
lower luminosities as summarized in Figures \ref{figME} and
\ref{figMN}.

SN 1997ef is found to be the HN class of energetic explosion, although
$E/M_{\rm ej}$ is a factor 3 smaller than GRB-SNe.  It is not clear
whether SN 1997ef is not associated with GRB because of this smaller
$E/M_{\rm ej}$ or it was actually associated with the candidate GRB
971115.

SN 2002ap was not associated a GRB and no radio has been observed.  It
has similar spectral features, but narrower and redder
(Fig. \ref{figSP}), which was modeled as a smaller energy explosion,
with $E_{51} \sim 4$ and $\Mej \sim 3
\Msun$\cite{maz02}.

The early time spectrum of SN 2003jd is similar to SN 2002ap.
Interestingly, its nebular spectrum shows a double peak in O-emission
lines \cite{maz05}.  This has exactly confirmed the theoretical
prediction by the asymmetric explosion model \cite{mae02}.  In this
case, the orientation effect might cause the non-detection of a GRB.

\subsubsection{XRF--Supernovae}

GRB060218 is the second closest event as ever ($\sim 140\,$Mpc).  The
GRB was weak \cite{campana2006} and classified as X-Ray Flash (XRF)
because of its soft spectrum.  The presence of SN 2006aj was soon
confirmed\cite{pian2006,mod06}.  Here we summarize the properties of
SN 2006aj by comparing with other SNe~Ic.

SN~2006aj has several features that make it unique.  It is less bright
than the other GRB/SNe (Fig. \ref{figSP}).  Its rapid photometric
evolution is very similar to that of a dimmer, non-GRB SN
2002ap\cite{maz02}, but it is somewhat faster.  Although its spectrum
is characterized by broader absorption lines as in SN 1998bw and other
GRB/SN, they are not as broad as those of SN~1998bw, and it is much
more similar to that of SN~2002ap (Fig. \ref{figSP}).  The most
interesting property of SN~2006aj is surprisingly weak oxygen lines,
much weaker than in Type Ic SNe.

By modeling the spectra and the light curve, we derive for SN~2006aj
$\Mej \sim 2 \Msun$ and $E_{51} \sim 2$.  Lack of oxygen in the
spectra indicates $\sim 1.3 \Msun$ of O, and oxygen is still the
dominant element.  We synthesize the theoretical light curve and find
that the best match is achieved with a total \Nifs\ mass of $0.21
\Msun$ in which $0.02 \Msun$ is located above 20,000\kms 
(Fig. \ref{figSP}).  

The properties of SN~2006aj (smaller $E$ and smaller $\Mej$) suggest
that SN~2006aj is not the same type of event as the other GRB-SNe
known thus far.  One possibility is that the initial mass of the
progenitor star is much smaller than the other GRB-SNe, so that the
collapse/explosion generated less energy.  If $M_{\rm ms}$ is $\sim 20
- 25 \Msun$, the star would be at the boundary between collapse to a
black hole or to a neutron star.  In this mass range, there are
indications of a spread in both $E$ and the mass of \Nifs\
synthesized\cite{hamuy2003}.  The fact that a relatively large amount
of \Nifs\ is required in SN 2006aj possibly suggests that the star
collapsed only to a neutron star because more core material would be
available to synthesize \Nifs\ in the case.

Although the kinetic energy of $E_{51}\sim 2$ is larger
than the canonical value ($1 \times 10^{51}$ erg, \cite{nom94}) in the
mass range of $M_{\rm ms} \sim 20 - 25 \Msun$, such an energy might be
obtained from magnetar-type activity.  

XRFs may be associated with less massive progenitor stars than those
of canonical GRBs, and that the two groups may be differentiated by
the formation of a neutron star\cite{nak98} or a BH.  In order for the
progenitor star to have been thoroughly stripped of its H and He
envelopes, the progenitor may be in a binary system.

\subsubsection{Non-SN Gamma-Ray Bursts}

For recently discovered nearby long-duration GRB~060505 ($z=0.089$,
\cite{fyn06}) and GRB~060614 ($z=0.125$, \cite{gal06,fyn06,del06,geh06}), 
no SN has been detected.  Upper limits to brightness of the possible
SNe are about 100 times fainter than SN~1998bw. These correspond to
upper limits to the ejected \Nifs\ mass of $\Mni\sim 10^{-3}\Msun$.

Tominaga et al. \cite{tom07} calculated the jet-induced explosions
(\eg \cite{mae03,nag06}) of the $40\Msun$ stars
\cite{ume05,tominaga2006} by injecting the jets at a radius 
$R \sim 900$ km, corresponding to an enclosed mass of $M \sim 1.4
\Msun$.  They investigated the dependence of nucleosynthesis outcome
on $\Ed$ for a range of $\Edep\equiv\Ed/10^{51}{\rm
ergs\,s^{-1}}=0.3-1500$.  The diversity of $\Ed$ is consistent with
the wide range of the observed isotropic equivalent $\gamma$-ray
energies and timescales of GRBs (\cite{ama06} and references therein).
Variations of activities of the central engines, possibly
corresponding to different rotational velocities or magnetic fields,
may well produce the variation of $\Ed$.

\begin{figure}[t]
\includegraphics[width=8.5cm]{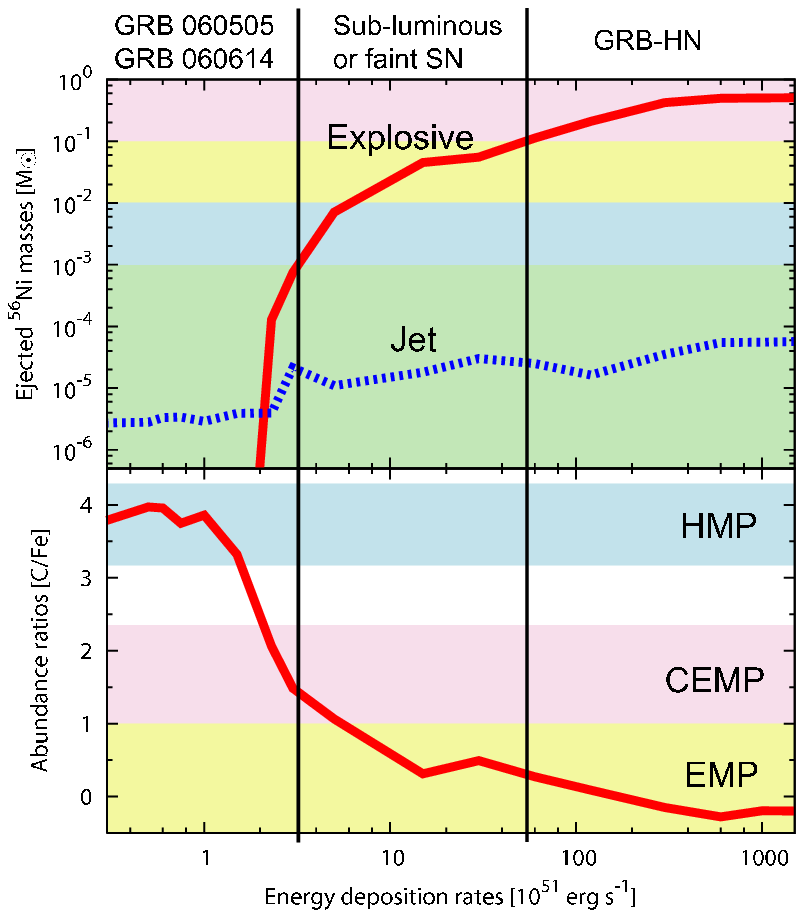}
\caption{{\it Top}: the ejected \Nifs\ mass ({\it red}: 
 explosive nucleosynthesis products, {\it blue}: the jet contribution)
 as a function of the energy deposition rate. The background color
 shows the corresponding SNe ({\it red}: GRB-HNe, {\it yellow}:
 sub-luminous SNe, {\it blue}: faint SNe, {\it green}: GRBs~060505 and
 060614).  Vertical lines divide the resulting SNe according to their
 brightness.  {\it Bottom}: the dependence of abundance ratio [C/Fe]
 on the energy deposition rate. The background color shows the
 corresponding metal-poor stars ({\it yellow}: EMP, {\it red}: CEMP,
 {\it blue}: HMP stars).
\label{fig:EdotNi}}
\end{figure}

\begin{figure}[t]
\includegraphics[width=10cm]{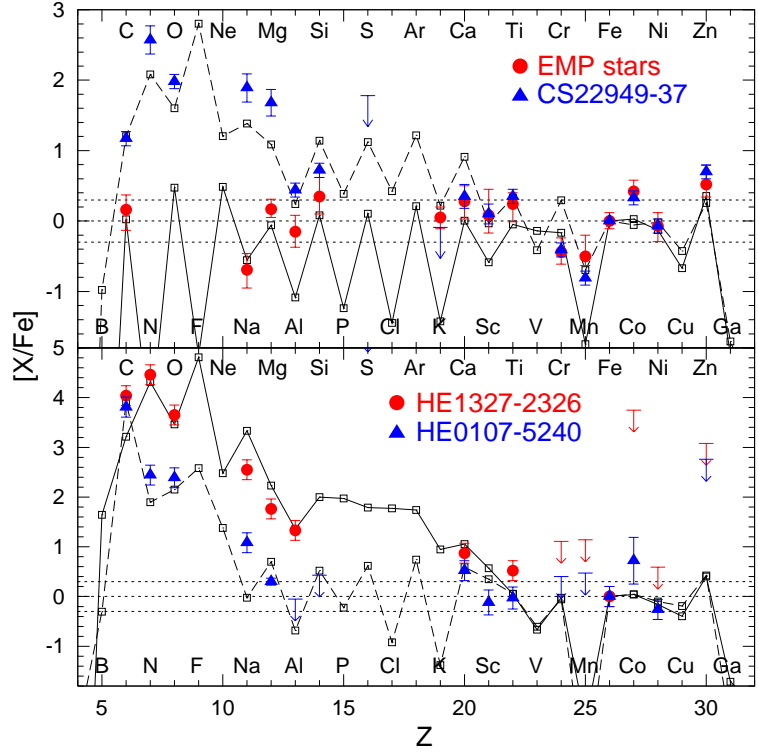}
\caption{
 A comparison of the abundance patterns of metal-poor 
 stars and of our models. 
 {\it Top}: typical EMP ({\it red dots}, \cite{cayrel2004}) and 
 CEMP ({\it blue triangles}, CS~22949--37, \cite{dep02}) stars and models
 with $\Edep=120$ ({\it solid line}) and $=3.0$ ({\it dashed line}).
 {\it Bottom}: HMP stars:
 HE~1327--2326, ({\it red dots}, \eg \cite{fre05}), 
 and HE~0107--5240, ({\it blue triangles}, \cite{chr02,bes05}) and models 
 with $\Edep=1.5$ ({\it solid line}) and $=0.5$ ({\it dashed line}).
\label{fig:HMP}}
\end{figure}

\section{Nucleosynthesis in Jet-Induced Explosions}

Nucleosynthetic properties found in the above diversity are connected
to the variation of the abundance patterns of extremely-metal-poor
stars, such as the excess of C, Co, Zn relative to Fe.  Such a
connection are modeled in a unified manner with the jet-induced
explosion model.

We have computed hydrodynamics and nucleosynthesis for the explosions
induced by relativistic jets. We have shown that (1) the explosions
with large energy deposition rate, $\Ed$, are observed as GRB-HNe and
their yields can explain the abundances of normal EMP stars, and (2)
the explosions with small $\Ed$ are observed as GRBs without bright
SNe and can be responsible for the formation of the CEMP and the HMP
stars.  We thus propose that GRB-HNe and GRBs without bright SNe
belong to a continuous series of BH-forming massive stellar deaths
with the relativistic jets of different $\Ed$.

\subsubsection{Diversity of \Nifs\ Mass}

The top panel of Figure~\ref{fig:EdotNi} shows the dependence of the
ejected $\Mni$ on the energy deposition rate $\Ed$. For lower $\Ed$,
smaller $\Mni$ is synthesized in explosive nucleosynthesis because of
lower post-shock densities and temperatures (\eg \cite{mae03,nag06}).

If $\Edep\gsim3$, the jet injection is initiated near the bottom of
the C+O layer, leading to the synthesis of $\Mni\gsim10^{-3}\Msun$. If
$\Edep<3$, on the other hand, the jet injection is delayed and
initiated near the surface of the C+O core; then the ejected \Nifs\ is
as small as $\Mni<10^{-3}\Msun$.

\Nifs\ contained in the relativistic jets is only
$\Mni\sim10^{-6}-10^{-4}\Msun$ because the total mass of the jets is
$M_{\rm jet}\sim 10^{-4}\Msun$ in our model with $\Gamma_{\rm
jet} = 100$ and $E_{\rm dep}=1.5\times10^{52}$ergs.  Thus the \Nifs\
production in the jets dominates over explosive nucleosynthesis in the
stellar mantle only for $\Edep<1.5$ in the present model.

For high energy deposition rates ($\Edep\gsim60$), the explosions
synthesize large $\Mni$ ($\gsim0.1\Msun$) being consistent with
GRB-HNe.  The remnant mass was $M_{\rm rem}^{\rm start}\sim1.5\Msun$
when the jet injection was started, but it grows as material is
accreted from the equatorial plane. The final BH masses range from
$M_{\rm BH}=10.8\Msun$ for $\Edep=60$ to $M_{\rm BH}=5.5\Msun$ for
$\Edep=1500$, which are consistent with the observed masses of
stellar-mass BHs \cite{bai98}.  The model with $\Edep=300$ synthesizes
$\Mni\sim0.4\Msun$ and the final mass of BH left after the explosion
is $M_{\rm BH}=6.4\Msun$.

For low energy deposition rates ($\Edep<3$), the ejected
\Nifs\ masses ($\Mni<10^{-3}\Msun$) are smaller than the upper
limits for GRBs~060505 and 060614. The final BH mass is larger for
smaller $\Ed$.
While the material ejected along the jet-direction involves those from
the C+O core, the material along the equatorial plane fall back.

If the explosion is viewed from the jet direction, we would observe
GRB without SN re-brightening. This may be the situation for
GRBs~060505 and 060614.  In particular, for $\Edep<1.5$, \Nifs\ cannot
be synthesized explosively and the jet component of the Fe-peak
elements dominates the total yields (Fig.~\ref{fig:HMP}). The
models eject very little $\Mni$ ($\sim10^{-6}\Msun$).

For intermediate energy deposition rates ($3\lsim\Edep<60$), the
explosions eject $10^{-3}\Msun \lsim \Mni <0.1\Msun$ and the final BH
masses are $10.8\Msun\lsim M_{\rm BH}< 15.1\Msun$. The resulting SN is
faint ($\Mni <0.01\Msun$) or sub-luminous ($0.01\Msun \lsim \Mni
<0.1\Msun$).

Nearby GRBs with faint or sub-luminous SNe have not been observed.
This may be because they do not occur intrinsically in our
neighborhood or because the number of observed cases is still too
small. In the latter case, further observations may detect GRBs with a
faint or sub-luminous SN.

\subsubsection{Abundance Patterns of C-rich Metal-Poor Stars}

The bottom panel of Figure~\ref{fig:EdotNi} shows the dependence of
the abundance ratio [C/Fe] on $\Ed$. Lower $\Ed$ yields larger $M_{\rm
BH}$ and thus larger [C/Fe], because the infall decreases the amount
of inner core material (Fe) relative to that of outer material (C)
(see also \cite{mae03}). As in the case of $\Mni$, [C/Fe] changes
dramatically at $\Edep\sim3$.

The abundance patterns of the EMP stars are good indicators of SN
nucleosynthesis because the Galaxy was effectively unmixed at [Fe/H]
$< -3$ (\eg \cite{tum06}). They are classified into three groups
according to [C/Fe]:

(1) [C/Fe] $\sim 0$, normal EMP stars ($-4<$ [Fe/H] $<-3$, \eg
    \cite{cayrel2004});

(2) [C/Fe] $\gsim+1$, Carbon-enhanced EMP (CEMP) stars ($-4<$ [Fe/H]
    $<-3$, \eg CS~22949--37 \cite{dep02}); 

(3) [C/Fe] $\sim +4$, hyper metal-poor (HMP) stars ([Fe/H] $<-5$, \eg
    HE~0107--5240 \cite{chr02,bes05}; HE~1327--2326 \cite{fre05}).

Figure \ref{fig:HMP} shows that the general abundance patterns of
the normal EMP stars, the CEMP star CS~22949--37, and the HMP stars
HE~0107--5240 and HE~1327--2326 are reproduced by models with
$\Edep=120$, 3.0, 1.5, and 0.5, respectively. The model for the normal
EMP stars ejects $\Mni\sim0.2\Msun$, i.e. a factor of 2 less than
SN~1998bw. On the other hand, the models for the CEMP and the HMP
stars eject $\Mni\sim8\times10^{-4}\Msun$ and $4\times 10^{-6}\Msun$,
respectively, which are always smaller than the upper limits for
GRBs~060505 and 060614.  The N/C ratio in the models for CS~22949--37
and HE~1327--2326 is enhanced by partial mixing between the He and H
layers during presupernova evolution \cite{iwamoto2005}.

\begin{figure}[t]
\includegraphics[width=10cm]{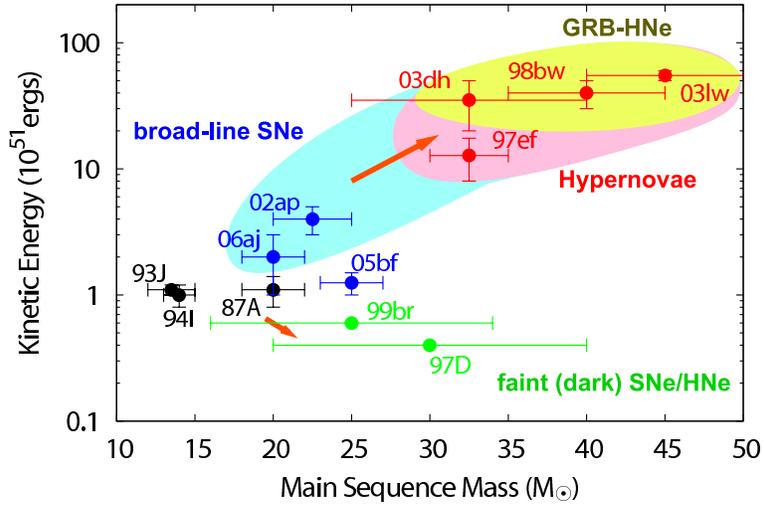}
\caption{
The kinetic explosion energy $E$ as a
function of the main sequence mass $M$ of the progenitors for several
supernovae/hypernovae. SNe that are observed to show broad-line features
are indicated.  Hypernovae are the SNe with $E_{51}>10$. 
}
\label{figME}
\end{figure}

\begin{figure}[t]
\includegraphics[width=10cm]{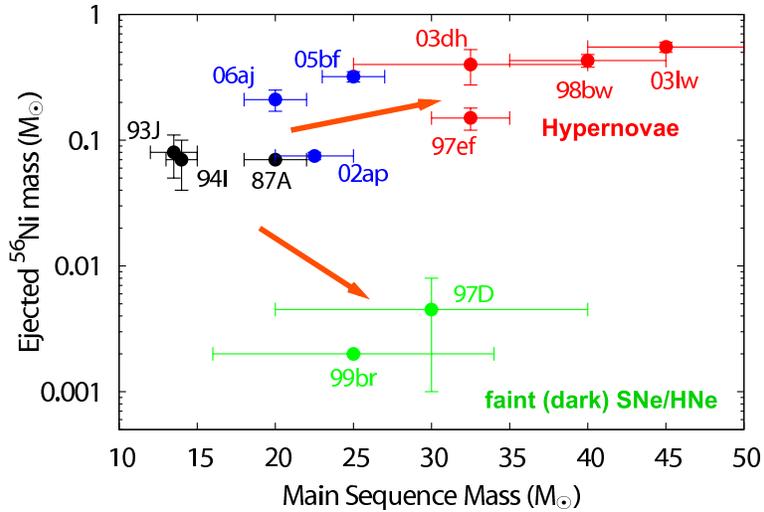}
\caption{
The ejected $^{56}$Ni mass as a
function of the main sequence mass $M$ of the progenitors for several
supernovae/hypernovae.  
}
\label{figMN}
\end{figure}

\section{Discussion}

The large Zn and Co abundances and the small Mn and Cr abundances
observed in very metal-poor stars can better be explained by
introducing HNe.  This would imply that HNe have made significant
contributions to the early Galactic chemical evolution,

In theoretical models, some element ratios, such as (K, Sc, Ti, V)/Fe,
are too small, while some ratios such as Cr/Fe are too large compared
with the observed abundance ratios \cite{cayrel2004}.  Underproduction
of Sc and K may require significantly higher entropy environment for
nucleosynthesis, e.g., the ``low density'' progenitor models for K,
Sc, and Ti \cite{ume05,fryer06}.  GRBs would have possible nucleosynthesis
site, such as accretion disks around the black hole \cite{prue05}.

Neutrino processes in the deepest layers of SN ejecta and a possible
accretion disk around a black hole would open a new window for SN
nucleosynthesis \cite{prue05, frohlich06, frohlich06b, wanajo2006}.

\subsubsection{GRB, Hypernovae, and Broad-Lines}

Figures \ref{figME} and \ref{figMN} summarize the properties of
core-collapse SNe as a function of the main-sequence mass $M_{\rm ms}$
of the progenitor star \cite{nomoto2003}.  
The broad-line SNe include both GRB-SNe and Non-GRB SNe.

(1) GRB vs. Non-GRB: Three GRB-SNe are all similar Hypernovae (i.e.,
$E_{51} \gsim$ 10.  Thus $E$ could be closely related to the formation
of GRBs.  SN 1997ef seems to be a marginal case.  $E/\Mej$ could be
more important because SN 1997ef has significantly smaller $E/\Mej$
than GRB-SNe.

(2) Broad-Line features: The mass contained at $v >$ 30,000 km
s$^{-1}$ (or even higher boundary velocity) might be critical in
forming the broad-line features, although further modeling is required
to clarify this point \cite{nomoto2006}.

\subsubsection{Black Holes vs. Neutron Stars}

The discovery of XRF~060218/SN~2006aj and their properties extend the
GRB-HN connection to XRFs and to the HN progenitor mass as low as
$\sim 20 \Msun$.  The XRF~060218 may be driven by a neutron star
rather than a black hole.

The final fate of 20 - 25 $\Msun$ stars show interesting variety.
Even normal SN Ib 2005bf is very different from previously known
SNe/HNe \cite{tom05b,fol05}.  This mass range corresponds to the
transition from the NS formation to the BH formation.  The NSs from
this mass range could be much more active than those from lower mass
range because of possibly much larger NS masses (near the maximum
mass) or possibly large magnetic field ({\em magnetar}).  XRFs and
GRBs from the mass range of 20 - 25 $\Msun$ might form a different
population.

\subsubsection{Hypernovae of Type II and Type Ib?}

Suppose that smaller losses of mass and angular momentum from low
metallicity massive stars lead to the formation of more rapidly
rotating NSs or BHs and thus more energetic explosions.  Then we
predict the existence of Type Ib and Type II HNe \cite{ham05}.

So far all observed HNe are of Type Ic.  However, most of SNe Ic are
suggested to have some He \cite{bra06}.  If even the small amount of
radioactive $^{56}$Ni is mixed in the He layer, the He feature should
be seen \cite{luc91,nom95}.  For HNe, the upper mass limit of He has
been estimated to be $\sim 2 \Msun$ \cite{maz02} for the case of no He
mixing.  If He features would be seen in future HN observations, it
would provide an important constraint on the models, especially, the
fully mixed WR models \cite{yoo05,woo06b,mey07}.

\begin{figure}[t]
\includegraphics[width=7.3cm]{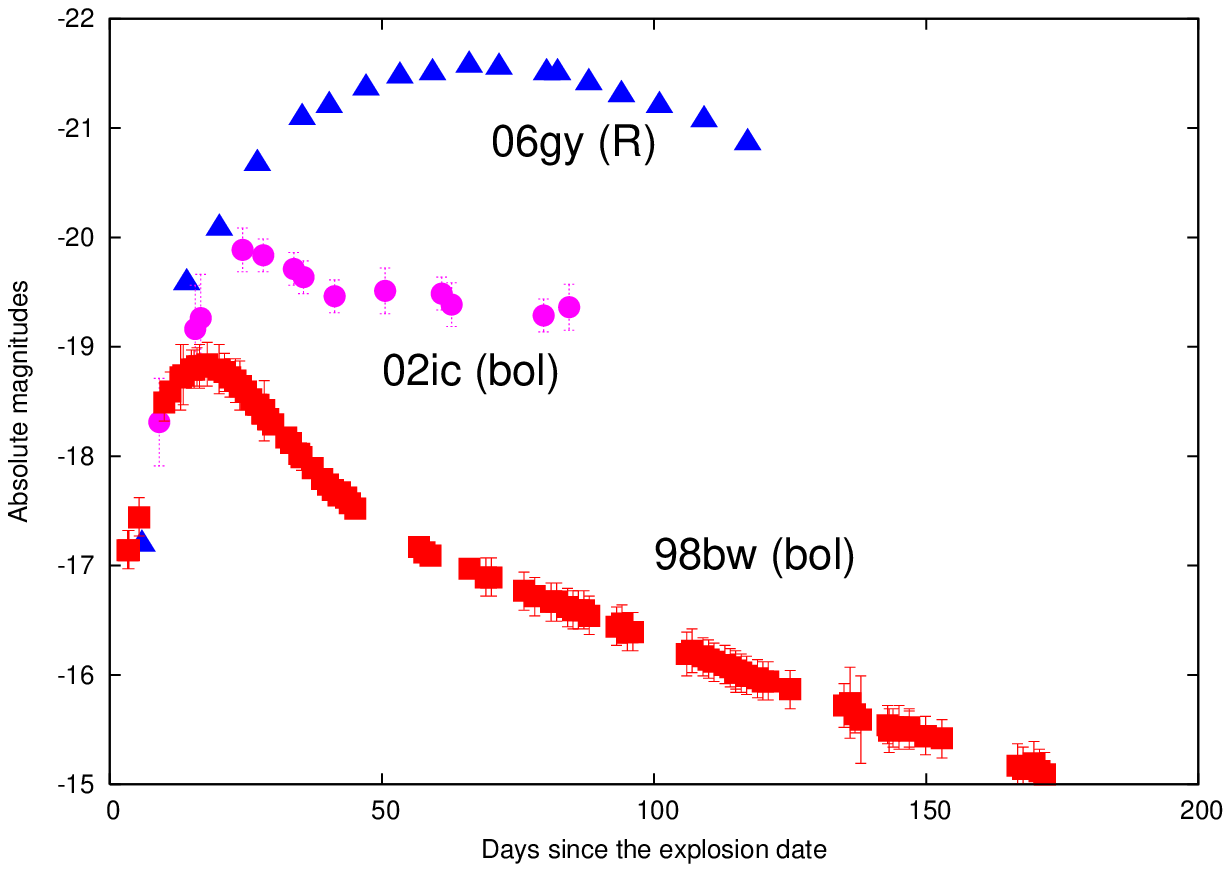}\\
\includegraphics[width=7.3cm]{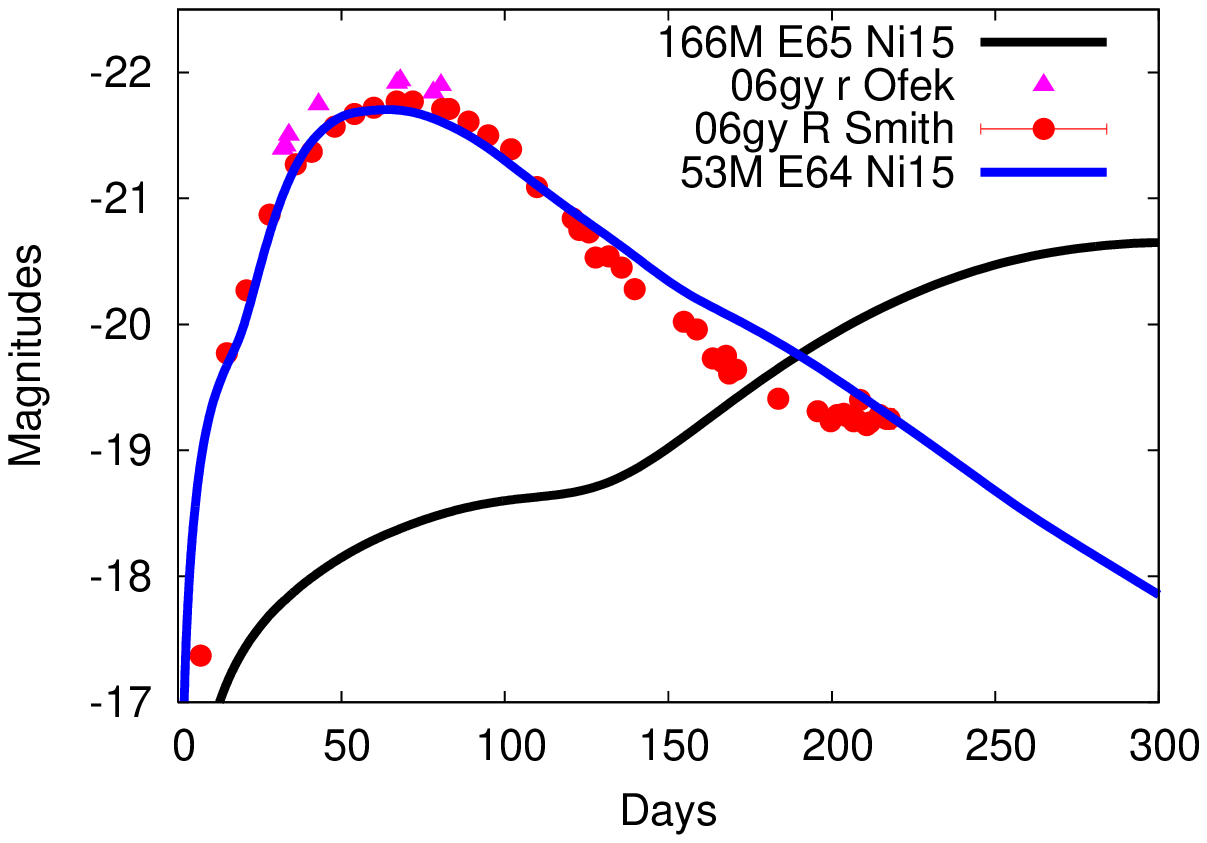}
\caption{(Left) Comparison between LCs of SNe~1998bw ({\it open
 square} \cite{galama1998}), 2002ic ({\it open circle} \cite{ham03}),
 and 2006gy ({\it fulled triangle} \cite{smi07}). (Right) Comparison
 between R- and r-band LCs of SN~2006gy \cite{ofe07,smi07} and
 synthetic LCs for a model with $\Mej=53\Msun$, $E_{51}=64$, and
 $M{\rm (^{56}Ni)}=15\Msun$ and a PISN model with $\Mej=166\Msun$,
 $E_{51}=65$, and $M{\rm (^{56}Ni)}=15\Msun$.}
\label{fig06gy}
\end{figure}

\section{Most Luminous Supernova 2006gy}

Recently, several extremely luminous supernovae have been discovered,
which includes SNe IIa 1997cy, 2002ic, and SN Ic 1999as
(Fig. \ref{fig06gy}: left).  The energy sources of these SN light
curves (LC) are closely related to SN nucleosynthesis.  The
post-maximum light curves (LCs) of SNe IIa are powered by
circumstellar interaction \cite{deng04}, although whether the
underlying SNe are Ia or Ic is under debate \cite{bene06}.

SN 2006gy is the most luminous SN \cite{ofe07,smi07}.  It shows
hydrogen emission features like SNe IIn and IIa to indicate
circumstellar interaction.  However, the X-ray luminosity is too low
to explain the observed optical luminosity \cite{smi07}.  This
suggests that the LC of SN 2006gy may be powered by the decays of
$^{56}$Ni $\to$ $^{56}$Co $\to$ $^{56}$Fe, and the required large
$^{56}$Ni mass, $M{\rm (^{56}Ni)}$, suggests that SN 2006gy could be a
pair-instability supernova (PISN) \cite{umeda2002a,heger02} rather
than a core-collapse \cite{ofe07,smi07}.

We have constructed the LC model of the PISN model as shown by the
solid line with 166M in Fig. \ref{fig06gy} (right).  Here we have
calculated the evolution from the main-sequence with extensive mass
loss to expose a C+O core.  The star undergoes PISN with
$\Mej=166\Msun$, $E_{51}=65$, and $M{\rm (^{56}Ni)}=15\Msun$.  Here
the explosion energy is not a free parameter (like a core-collapse
model) but obtained from the nuclear energy release associated with
the production of $M{\rm (^{56}Ni)}$.  Such a large $\Mej$, which is
necessary to produce large enough $M{\rm (^{56}Ni)}$, is too large for
the LC.  The model LC evolves much {\em slower} than the observed LC
of SN 2006gy (red points in Fig. \ref{fig06gy}: right).

In order to reproduce the LC of SN 2006gy, we artificially reduce the
ejected mass of the above exploding model down to $\Mej=53\Msun$,
keeping $E_{51}=64$, and $M{\rm (^{56}Ni)}=15\Msun$ (the 53M line in
Fig. \ref{fig06gy}).  In other words, the progenitor should have lost
much more mass than the actual model, yet produced a large enough
amount of $^{56}$Ni.

These results imply whether SN 2006gy is a core-collapse SN or a PISN
is not clear yet.  The PISN model can produce enough $M{\rm
(^{56}Ni)}$ but $\Mej$ might be too large.  The core-collapse models
currently available could produce too small $M{\rm (^{56}Ni)}$.

Also, the mass loss from such a massive star with solar metallicity
would be too large to keep hydrogen-rich circumstellar matter as
observed like SNe IIn and IIa \cite{ofe07}.  Stellar merging in a
close binary system or a dense star cluster might be necessary for the
formation and evolution of very massive stars.  Further study is
clearly necessary to understand the evolutionary origin and
nucleosynthesis of SN 2006gy.

%%%%%%%%%%%%%%%%%%%%%%%%%%%%%%%%%%%%%%%%%%%%%%%%
%% BACKMATTER
%%%%%%%%%%%%%%%%%%%%%%%%%%%%%%%%%%%%%%%%%%%%%%%%

%\begin{theacknowledgments}
%This work has been supported in part by the Grant-in-Aid for
%Scientific Research (17030005, 17033002, 18104003, 18540231 for K.N.)
%and the 21st Century COE Program (QUEST) from the JSPS and MEXT of
%Japan.
%\end{theacknowledgments}



\bibliographystyle{aipproc}   % if natbib is available

\bibliography{sample}

%  \typeout{******************************************}
%  \typeout{** Please run "bibtex \jobname" to optain}
%  \typeout{** the bibliography and then re-run LaTeX}
%  \typeout{** twice to fix the references!}
%  \typeout{******************************************}

%\end{document}

%%%%%%%%%%%%%%%%%%%%%%%%%%%%%%%%%%%%%%%%%%%
%% The following lines show an example how to produce a bibliography
%% without the help of the BibTeX program. This could be used instead
%% of the above.
%%%%%%%%%%%%%%%%%%%%%%%%%%%%%%%%%%%%%%%%%%%

%\endinput
\end{document}